**Thermoelectric power factor enhancement by spin-polarized currents – a nanowire case study**

*Anna Corinna Niemann,\* Tim Böhnert, Ann-Kathrin Michel, Svenja Bäßler, Bernd Gotsmann, Katalin Neuróhr, Bence Tóth, László Péter, Imre Bakonyi, Victor Vega, Victor M. Prida, Johannes Gooth, and Kornelius Nielsch\**

A.C. Niemann, A.-K. Michel, S.Bäßler, Dr. J. Gooth, Prof. Dr. K. Nielsch
Institute of Nanostructure and Solid State Physics, Universität Hamburg, Jungiusstraße 11, D-20355, Hamburg, Germany
E-mail: aniemann@physnet.uni-hamburg.de

Dr. T. Böhnert
International Iberian Nanotechnology Laboratory, Av. Mestre Jose Veiga, Braga 4715, Portugal

Dr. J. Gooth, Dr. B. Gotsmann
IBM Research-Zurich, Säumerstrasse 4, 8803 Rüschlikon, Switzerland

Dr. K. Neuróhr, Dr. B.G. Tóth, Prof. Dr. L. Péter, Prof. Dr. I. Bakonyi
Wigner Research Center of Physics, Hungarian Academy of Sciences,
Konkoly-Thege út 29-33, H-1121, Budapest, Hungary

Dr. V. Vega, Prof. Dr. V. M. Prida
Depto. Física, Universidad de Oviedo, E-33006, Oviedo, Spain

Prof. Dr. K. Nielsch
Leibniz Institute for Solid State and Material Research, Helmholtzstraße 20,
D-01069, Dresden, Germany
E-mail:k.nielsch@ifw-dresden.de

Keywords: electrodeposition, nanowires, magnetoresistance, nanostructured thermoelectrics

Thermoelectric (TE) measurements have been performed on the workhorses of today's data storage devices, exhibiting either the giant or the anisotropic magnetoresistance effect (GMR and AMR). The temperature-dependent (50–300 K) and magnetic field-dependent (up to 1 T) TE power factor (PF) has been determined for several Co-Ni alloy nanowires with varying Co:Ni ratios as well as for Co-Ni/Cu multilayered nanowires with various Cu layer thicknesses, which were all synthesized via a template-assisted electrodeposition process. A systematic investigation of the resistivity, as well as the Seebeck coefficient, is performed for

Co-Ni alloy nanowires and Co-Ni/Cu multilayered nanowires. At room temperature, measured values of TE PFs up to 3.6 mWK$^{-2}$m$^{-1}$ for AMR samples and 2.0 mWK$^{-2}$m$^{-1}$ for GMR nanowires are obtained. Furthermore, the TE PF is found to increase by up to 13.1 % for AMR Co-Ni alloy nanowires and by up to 52 % for GMR Co-Ni/Cu samples in an external applied magnetic field. The magnetic nanowires exhibit TE PFs that are of the same order of magnitude as TE PFs of Bi-Sb-Se-Te based thermoelectric materials and, additionally, give the opportunity to adjust the TE power output to changing loads and hotspots through external magnetic fields.

1. **Introduction:**

In recent years, the continuously increasing integration density of microelectronic devices has brought thermoelectric (TE) effects back into the focus of current research. Taking advantage of the nanostructuring capabilities of semiconductors, thermoelectric energy converters exploiting the Seebeck effect have been designed with the aim of recapturing the energy wasted by heat loss and converting it to electricity.[1] Furthermore, scaling towards lower device currents is concurrent with miniaturization; and, consequently, Peltier heating is becoming significant or even dominant over Joule heating due to an increasing interface density in microelectronic devices.[2] In addition, large temperature gradients change device currents through the Thomson effect.[3] Nanostructuring of thermoelectric materials has also been recently proposed as a novel route for properly tailoring the parameters of the figure of merit coefficient, ZT, due to the opportunity of tuning the various appropriated features of the materials at the nanoscale. Among others, downscaling may open the way to decouple the thermal and electrical conductivity together with a reduction of the thermal conductivity by increasing the scattering of phonons through interfaces or defects.[4, 5]

Whether or not TE effects are advantageous or problematic for respective device performances depends on the specific application. In phase-change memory devices, for example, local heating is exploited as the mechanism for switching the resistance state of a nanoscale memristor,[6] Peltier effects modify localized hotspots in logic devices,[7] and local temperature gradients imposed within device structures can reach extreme values leading to Thomson and Peltier effects that can reach high magnitudes. By neglecting them, one may fail to predict device switching properties by more than 40%.[3]

An interesting class of integrated circuit elements is based on magnetoresistance (MR) effects, which nowadays provide an integral concept in omnipresent electronic devices such as reading heads for magnetic hard discs, magnetic sensors and magnetoresistive random-access memories (MRAMs).[8] Both anisotropic magnetoresistance (AMR)[9-11] and giant magnetoresistance (GMR)[12, 13] effects that are based on spin-dependent transport in magnetic alloys and magnetic/nonmagnetic multilayers have found extensive application. Driven by the need for constantly increasing data storage densities and the corresponding enhancement of magnetoresistance sensitivity, tremendous progress in the miniaturization of MR devices has been made by down-sizing toward the nanometer scale.

Recently, TE effects have also been considered in spintronic devices[14] and spin-dependent Seebeck effects were observed for various MR regimes, such as AMR,[15] GMR[16] as well as tunnelling MR.[17] Furthermore, macroscopic TE signals have been linked to nanoscopic spin configurations of domain walls in ferromagnetic nanostructures.[18]

In the present work, we study and compare the thermoelectric properties of standard magnetic materials in the Co-Ni-Cu systems. We chose two model systems: Co-Ni alloy nanowires representing the AMR regime on the one hand and Co-Ni/Cu multilayered nanowires representing the GMR regime on the other hand. Both have well-defined magnetization directions along the nanowire axis and in the plane of the magnetic layers, respectively. Not only the thermoelectric power (also called Seebeck coefficient $S$), but also its dependence on the external applied magnetic field, was investigated. Large TE power factors (PF = $S^2/\rho$, with $\rho$ being the electrical resistivity) of up to 4 mWK$^{-2}$m$^{-1}$ are observed, which can easily compete with common TE materials used specifically for power generation. Moreover, TE performances were modified by up to 52% of the zero-field values by applying an external magnetic field, allowing for adjustability to changing loads and hotspots. In the framework of the two-current model,[19] we show that this increase of performance owes to the fact that the spin channel, exhibiting a lower $\rho$, simultaneously has a higher $S$ than the other spin channel, as depicted in **Figure 1.**

2. **Results and Discussion:**

**2.1 Synthesis, Compositional and Structural Characterization:**

The magnetic nanowires were synthesized by template-assisted electrodeposition into self-ordered, hard-anodized alumina (HAAO) membranes[20, 21] with the internal walls of the membranes coated with SiO$_2$ in an atomic layer deposition (ALD) process in order to enhance the physical stability of the future nanowires and prevent them from oxidation after dissolving the alumina templates.[22, 23] Alloy and multilayered nanowires were deposited by procedures which were described in detail by Vega *et al.*[24] and Tóth *et al.*.[25] After the deposition process, we dissolved the HAAO templates in chromic-phosphoric acid aqueous solution and suspended the individual wires in ethanol.

The geometrical and compositional parameters of the alloy and multilayered nanowires investigated in the present work are summarized in **Table 1**.

Individual alloy nanowires were investigated by transmission electron microscopy (TEM), and their chemical composition was analyzed by electron dispersive x-ray spectroscopy (EDS) with the TEM, as shown in **Figure 2** (a) and (b), respectively. In total, one Ni sample and three different Co-Ni alloy samples with Co content ranging from 24 % up to 71 % were prepared, as listed in Table 1 (a).

The crystalline phase of the Co-Ni alloy wires was characterized by x-ray diffraction (XRD).[24] Ni-rich nanowires (Ni, $Co_{24}Ni_{76}$, and $Co_{39}Ni_{61}$) exhibit a face centered cubic (fcc) structure, whereas in the Co-richest sample ($Co_{71}Ni_{29}$) the hexagonally close packed (hcp) phase and the fcc phase coexist.[24] The crystallite size of the nanowires obtained from XRD line width analysis[24] was 5±2 nm for fcc crystals and 9±2 nm for hcp crystals. The XRD results obtained on the crystal structure with the nanowires inside the membrane were supported by TEM selected-area electron diffraction (TEM-SAED) patterns on single fcc and hcp nanowires[26] as shown in the inset of Figure 2 (a) for the hcp nanowire sample $Co_{71}Ni_{29}$.

In the same way, self-standing Co-Ni/Cu multilayered nanowires were analyzed by TEM and TEM-EDS,[27] as exemplified in Figure 2 (c) and (d), respectively. Bilayer thicknesses were measured over ten Co-Ni/Cu layers from high-resolution TEM (HR-TEM) images and the chemical composition of the nanowires was measured by TEM-EDS line scans performed on single nanowires, as given in Table 1 (b). From the EDS line scan, a counter oscillation of the Cu content and the amount of Co and Ni can be observed. The Co:Ni ratio in the magnetic layers of the six different multilayered nanowire samples ranges from 50:50 to 30:70. Also, a Cu content of about 2% can be estimated in the magnetic segments,[25] which is a value well below the absolute uncertainty of the chemical composition analysis of about 5%. From the bilayer thicknesses and the chemical composition, Cu layer thicknesses, ranging from 0.2 nm up to 5.2 nm, and Co-Ni layer thicknesses from 5.2 nm up to 17.4 nm were calculated, as shown in Table 1(b).

In the following, alloy nanowires will be referred to according to their chemical composition – Ni and Co-Ni – and multilayered wires will be named according to their Cu spacer layer thickness Cu (x nm), since these are the main defining parameters of the anisotropic and giant magnetoresistance regime, respectively.

By using SEM, we determined the length between the electrical contacts, $\ell$, which was about 8 μm for all wires as well as the diameter, $d$, of the nanowires, which varied between 126 nm and 258 nm, as listed in Table 1(a) and (b). To calculate the resistivity $\rho = R\pi d^2/4\ell$ of the nanowires with $R$ being the resistance, we have also corrected the measured diameters for a 5 nm-thick $SiO_2$ shell, since it does not contribute to the electrical transport.

## 2.2 Thermoelectric Characterization:

Individual nanowires dispersed on a glass substrate were electrically contacted by defining a photoresist mask by laser beam lithography and subsequent sputtering of a Ti adhesion and a Pt layer. The inner part of the microstructure with an embedded nanowire can be seen on a SEM image of **Figure 3** (a).

Transport measurements were conducted in a physical property measurement system (PPMS) at different base temperatures, $T$, from 50 K to 325 K and at an applied magnetic field, $\mu_0 H_\perp$, perpendicular to the nanowire axis. To the heater line of the microdevice, DC voltages between 5 V and 14 V were applied, which caused Joule heating between 1 mW and 8 mW. On a typical microdevice, this led to temperature differences between 5 K and 10 K between the two resistive thermometers. The thermovoltage $U_{thermo}$ was recorded as a dc voltage and $R$ of the nanowire as well as of the two resistive thermometers were measured by employing standard lock-in techniques.

Plotting $R$ as a function of $\mu_0 H_\perp$ gives typical magnetoresistance curves, as illustratively shown in Figure 3(b) for one GMR-sample of Cu (5.2 nm) and one AMR-nanowire of Ni (data obtained at room temperature). Magnetoresistance was calculated as MR = $(R_H-R_0)/R_0$, the resistance $R_0$ being obtained without an external magnetic field and the resistance $R_H$ measured at a certain magnetic field $H$. While the Ni nanowire exhibits relatively small MR of -2.0 % at the magnetization saturation field value; the multilayered Cu (5.2 nm) nanowire exhibits larger MR ratio of up to -11.0 %, as shown in Figure 3(c).

An enhanced absolute thermopower $S = U_{thermo}/(T_{hot}-T_{cold})$ was measured by increasing $\mu_0H_\perp$ for alloy as well as multilayered nanowires, see Figure 3(d). The observation that $S$ increases and $\rho$ decreases with increasing $\mu_0H_\perp$ is in good agreement with several publications.[16] [28-32] We quantified the thermopower change with $\mu_0H_\perp$ by the magneto-thermoelectric power MTEP = $(S_H-S_0)/S_0$, $S_0$ and $S_H$ being the thermopower without an external magnetic field and at a certain applied magnetic field $H$, respectively.[33] Again, the Ni nanowire exhibits a relatively low magneto-thermoelectric power effect of MTEP = 4.6 % at the magnetization saturation field, and the multilayered Cu (5.2 nm) nanowire has a larger MTEP of up to 14.3 %, as shown in Figure 3(e). Comparing MR and MTEP of all the alloy and multilayered nanowires investigated, the following key observations[26, 27] have been made: First, MR of the alloy wires reaches bulk-like literature values,[10, 11] [34] except for $Co_{71}Ni_{29}$, which can be attributed to the strong magneto-crystalline anisotropy of the sample and the therefore non-negligible MR component in a magnetic field applied parallel to the nanowire axis. Secondly, at room temperature, multilayered wires with sufficiently thick Cu spacer layers show MR values from -6 % up to -15 %, comparable to the ones reported in other publications,[35-43] while wires with a Cu layer thickness below 1 nm show significantly reduced MR, ranging from -3.3 % to -3.6 %. This low performance can be attributed to pinholes in the nonmagnetic spacer layer and the resulting inability of the neighbouring magnetic layers to establish an antiparallel (or, at least, random or non-parallel) magnetization alignment in zero magnetic field;[44] Thirdly, the thermopower change under an external applied magnetic field of all alloy and multilayered wires is in the same range as the corresponding MR effects of the samples. An in-depth discussion of MR and MTEP can be found in the recently published work by Böhnert *et al.*.[26, 27]

The resistivity, $\rho$, of alloy[26] and multilayered[27] samples exhibits a temperature dependence which is characteristic of metals in general. At room temperature, $\rho$ of the alloy wires ranging from $\rho(Ni) = 13.2\ \mu\Omega cm$ to $\rho(Co_{24}Ni_{76}) = 21.7\ \mu\Omega cm$ is higher than the resistivity of the corresponding compositions in the bulk Ni-Co alloy system.[45] This deviation is well known in literature and is attributed to the nanocrystalline structure of electrodeposited materials.[45] At the same time, $\rho$ of the multilayered nanowires, ranging from $\rho(Cu\ (5.2\ nm)) = 28.7\ \mu\Omega cm$ to $\rho(Cu\ (1.4\ nm)) = 50.8\ \mu\Omega cm$, is further enhanced when compared to $\rho$ of the alloy wires, which we attribute to an additional scattering at the Co-Ni/Cu interfaces. The high uncertainty of the measured diameter and length of the nanowires leads to a high uncertainty of $\rho$ up to 15 %.

The resistivity and the absolute value of $S$ show a similar increase with the mean sample temperature $\bar{T} = (T_{hot} + T_{cold})/2$ for both alloy[26] and multilayered[27] nanowires. $S(300\ K)$ of the alloy wires ranges between -15.6 $\mu VK^{-1}$ (Ni) and -26.5 $\mu VK^{-1}$ ($Co_{71}Ni_{29}$) and $S(300\ K)$ of the multilayered samples was measured between -10.5 $\mu VK^{-1}$ (Cu (3.5 nm)) and -24.5 $\mu VK^{-1}$ (Cu (0.2 nm)). Consequently, a local hotspot-induced temperature difference of 100 K along an MRAM device would cause local potential variations in the mV range. In fact, also the reverse effect, namely the Peltier effect describing a temperature gradient across an interface resulting from a voltage drop may have a significant influence on the MR device performance, being particularly relevant for GMR devices. By using the Thomson relation, we have calculated the Peltier coefficient $\Pi = S \cdot \bar{T}$ of our samples to range from -3.15 mV (Cu (3.5 nm)) to -7.95 mV ($Co_{71}Ni_{29}$) at room temperature. Thus, a Peltier heat flow $\dot{Q} = \Pi \cdot I$ of the $Co_{71}Ni_{29}$ nanowire is generated, which is enhanced by as much as a factor of 8.0 ($R(Co_{71}Ni_{29}) = 100\ \Omega$) as compared to the heat generated by Joule heating when using a measurement current of 10 $\mu A$, and which, therefore, should not be neglected. One has to keep in mind that both $S$ and $\Pi$ coefficients refer not only to the nanowires but also to the other microdevice material Pt, Au, and Pt-Cr as listed in the legend of **Figure 4**.

The TE power output of the investigated nanowires, which is given by PF shown in Figure 4(a) and (b), increases monotonically with $\bar{T}$ for all samples, except for $Co_{24}Ni_{76}$ and Cu (3.5 nm). At room temperature, the PF of the alloy samples ranges from 1.9 mWK$^{-2}$m$^{-1}$ for Ni to 3.6 mWK$^{-2}$m$^{-1}$ for the $Co_{71}Ni_{29}$ nanowire. In the case of the multilayered samples, the PF(300 K) ranges from 0.3 mWK$^{-2}$m$^{-1}$ for Cu (3.5 nm) to 2.0 mWK$^{-2}$m$^{-1}$ for Cu (0.2 nm). The TE power output of these magnetic nanowires can compete with PFs of the best TE bulk semiconductor materials like $Bi_2Te_3$,[46] which gives a PF of 1.9 mW/K$^{-2}$m$^{-1}$ at room temperature and it exceeds the PF of $Bi_2Te_3$ nanowires by a factor of five.[47] Thus, even if the magnetic metallic nanowires cannot reach values of $S$ as high as the classical TE material ($S(Bi_2Te_{3,bulk}) = $ -168 μVK$^{-1}$),[46] the resistivity of the latter is much higher ($\rho(Bi_2Te_{3,bulk}) = $ 1430 μΩcm),[46] and this allows the magnetic metallic nanowires to easily compete with and even exceed the TE power output of Bi-Se-Sb-Te material systems. Therefore, these materials may be interesting candidates for TE power generation under specific environments as well as heat management in micro- and nano-scale electronic devices.

The thermoelectric power output at – or slightly above – room temperature is probably most relevant for heat dissipation and device applications. We have recorded the magnetic field-dependence of the PF at 300 K as shown in **Figure 5** (a) and (c) for alloy and multilayered nanowires, respectively. It can be observed that the TE power output of all samples increases in an applied magnetic field until the magnetization saturation field is reached. Alloy samples reach PFs of up to 4.2 mWK$^{-2}$m$^{-1}$ ($Co_{71}Ni_{29}$) and multilayered nanowires show PFs of up to 2.2 mWK$^{-2}$m$^{-1}$ (Cu (0.2 nm)) at the magnetization saturation field. To quantify the change of PF in an applied magnetic field, we define the magneto-PF ratio MPF= (PF$_H$-PF$_0$)/PF$_0$, with PF$_0$ the TE PF without an external applied magnetic field and PF$_H$ the TE PF at a certain magnetic field applied perpendicular to the nanowire axis. The MPF is shown in Figure 5(b) and (d) as a function of $\mu_0 H_\perp$ for alloy and multilayered nanowires, respectively. For alloy wires, the MPF ranges from 4.9 % ($Co_{71}Ni_{29}$) up to 13.1 % ($Co_{39}Ni_{61}$), while for multilayered wires an increase of the TE power output from 9.8 % (Cu (0.2 nm)) to as high as 52.2 % (Cu (3.5 nm)) is seen. For both, the AMR and GMR systems, the improved TE performance in an increasing applied magnetic field can be understood within the framework the two-current model,[19] which in this case describes two conducting spin-channels in a parallel circuit, as depicted in Figure 1. The TE PF of AMR as well as GMR systems benefits from the fact that for a high degree of spin-polarization, the prevailing low resistive majority spin channel simultaneously exhibits the highest thermopower within the system. To understand this, consider two conducting spin channels, the majority, ↑, and minority, ↓, channel. The total resistivity of the system is then given by $\rho_{tot} = (1/\rho_\uparrow + 1/\rho_\downarrow)^{-1}$ and it is well known that in high external magnetic fields the spin channel with the lower resistivity prevails. We assume that $\rho_\uparrow \ll \rho_\downarrow$, so that in the limit of full spin polarization $\rho_{tot}(\mu_0 H \geq \mu_0 H_{sat}) = \rho_\uparrow > \rho_{tot}(\mu_0 H = 0\,\text{T})$. The corresponding total thermopower of the system[48, 49] is given by $S_{tot} = (S_\uparrow/\rho_\uparrow + S_\downarrow/\rho_\downarrow) / (1/\rho_\uparrow + 1/\rho_\downarrow)$, where $min[S_\uparrow, S_\downarrow] < S_{tot} < max[S_\uparrow, S_\downarrow]$. In large applied magnetic fields, it follows directly from $\rho_\uparrow \ll \rho_\downarrow$ that $S_{tot} \approx (S_\uparrow/\rho_\uparrow) / (1/\rho_\uparrow) = S_\uparrow$. Since in our experiments the thermopower increases with an increasing external magnetic field, $S_{tot} = max[S_\uparrow, S_\downarrow]$ and $S_\uparrow \gg S_\downarrow$ follows. Consequently, the thermoelectric power for the AMR as well as for the GMR samples is maximized for highest spin polarization.

We note that regarding heat management in nanostructured devices, also the thermal conductivity, $\kappa$, is of major interest. Commonly, the Wiedemann-Franz law $\kappa \cdot \rho = L \cdot \bar{T}$ is used to calculate $\kappa$ from $\rho$, with the Lorenz number $L$ being the Sommerfeld value $L_0 = 2.45 \cdot 10^{-8}\,\text{V}^2\text{K}^{-2}$ for bulk metals above the Debye temperature.[50] These considerations lead to $\kappa(300\,\text{K})$ values from 14.41 $\text{Wm}^{-1}\text{K}^{-1}$ (Cu (1.4 nm)) up to 55.45 $\text{Wm}^{-1}\text{K}^{-1}$ (Ni) for the metallic magnetic nanowires investigated here. Furthermore, the thermoelectric figure of merit ZT ($ZT = S^2T/\rho\kappa = S^2/L_0$) ranges from 0.005 up to 0.009 at room temperature. However, in-plane measurements of $\rho$ and $\kappa$ on Co/Cu multilayered thin films[51, 52] and Ni nanowires[53] have recently shown that their $L$ value deviates from $L_0$ and depends on the magnetic configuration. Consequently, a calculation of $\kappa$ and ZT for our nanowires using the Wiedemann-Franz law combined with $L_0$ is only a rough estimation and measuring both $\rho$ and $\kappa$ of magnetic nanostructures remains an interesting challenge and highly anticipates further studies in the near future.

## 3. Conclusions:

To address emerging challenges and opportunities caused by heat management at the micro- and nanoscale, we studied the TE performance of magnetic nanowires for two different magnetoresistance regimes, namely anisotropic magnetoresistance and giant magnetoresistance.

At room temperature, for the AMR nanowires PFs of up to 3.6 mWK$^{-2}$m$^{-1}$ (Co$_{71}$Ni$_{29}$) and for the GMR wires PFs of up to 2.0 mWK$^{-2}$m$^{-1}$ (Cu (0.2 nm)) have been found. The TE PF of both alloy and multilayered nanowires increases with increasing mean temperature of the nanowires. Thus, for heat dissipating structures which might be operated above room temperature, even higher TE power outputs can be expected. Our data show that TE power outputs of magnetic nanowires can compete with the power outputs of classical TE bulk semiconductor materials like Bi$_2$Te$_3$[46] and clearly exceed the PF of nanostructured thermoelectric Bi$_2$Te$_3$ nanowires.[47] Additionally, the metallic nanowires can also be expected to exhibit a higher thermal conductivity than Bi-Sb-Te-Se material systems. By applying an external magnetic field, the TE power outputs of the AMR as well as of the GRM nanowires can be further increased, due to the unique property of such systems that the dominating majority spin channel simultaneously exhibits lowest resistivity and highest thermopower in the system. At room temperature, the largest enhancement of PF in an applied magnetization saturation field was measured to be 13.1 % for $Co_{39}Ni_{61}$ alloy AMR wires and 52 % for Cu (3.5 nm) multilayered GMR wires.

Therefore, we propose that MR nanodevices are interesting for heat dissipating applications due to the adjustability of their transport properties to changing loads and local hotspots by an applied magnetic field in nano- and microscale electronics. Furthermore, we showed that TE power generation from waste heat on the nanoscale is becoming into a competitive scenario compared to conventional thermoelectric materials since the magnetic nanowires can achieve equally high or even higher PFs than the semiconductor TE materials. Finally, this study also revealed the importance of taking significant TE effects into account when designing MR operational units.

## 4. Experimental section:

*Nanowire synthesis:* The nanowires were electrodeposited into self-ordered hard anodized alumina (HAAO) membranes which exhibit a nanoporous structure with a pore diameter between 136 nm and 268 nm. Before the anodization process, cleaned and electropolished high-purity aluminum foils (Al 99.999 %) were pre-anodized under mild anodization conditions ($U = 80$ V, $t = 10$ min) in an 0.3 M oxalic acid aqueous solution containing 5 vol. % of ethanol. We increased the applied voltage by 0.08 V s$^{-1}$ to perform the hard anodization process at 140 V ($t = 1.5$ h, 0 °C < $T$ < 3 °C). After anodization, the Al backside of the AAO membranes was removed by wet chemical etching in $CuCl_2$/HCl aqueous solution. Subsequently, the alumina bottom layer of the membrane pore structure as well as the protective mild anodized layer at the top side of the membrane was opened by immersing the membrane in a 5 wt. % $H_3PO_4$ solution ($t = 2.5$ h, $T = 30$ °C). Using atomic layer deposition (ALD) technique, the porous structure of the alumina membranes was homogeneously covered with a 5 nm thick $SiO_2$ layer. Reactive ion etching removes the $SiO_2$ layer at the top and backsides of the membrane. The backside of the AAO membrane was then again sealed by sputtering and electrodeposition of a gold layer. The deposition process of the nanowires took place in a three-electrode setup with the gold layer at the membrane backside acting as the working electrode, a platinum mesh serving as the counter electrode, and an Ag/AgCl (3 M) KCl reference electrode. Alloy nanowires with a length of 15 µm – 30 µm were electrodeposited at 35 °C and under potentiostatic conditions (-0.8 V > $U$ > -1 V vs. reference electrode) from a Watts-type electrolyte containing different compositions of $Co^{2+}$ and $Ni^{2+}$ ions as described by Vega et al..[24] The multilayered nanowires were grown by two-pulse plating from a single electrolyte according to a recipe published by Tóth et al..[25] Co-Ni and Cu layers were deposited at -1.5 V and -0.58 V vs. reference electrode, respectively. The desired Co-Ni and Cu layer thicknesses were achieved by properly adjusting the deposited charge equivalent. After the deposition process, the gold electrode was removed from the backside of the HAAO membrane by using (0.6 M) KI · (0.1 M) $I_2$ aqueous solution and then the membranes were dissolved in chromic-phosphoric acid aqueous solution (1.8 wt. % $CrO_3$, 6 wt. % $H_3PO_4$) for approximately 48 hours at 45 °C. The released nanowires were first filtrated and rinsed with deionized water and afterwards stored in ethanol.

*Structural and compositional characterization:* The morphology and chemical composition of the alloy nanowire arrays inside the membrane were characterized using a scanning electron microscope (Supra 55-Zeiss) equipped with EDS. The crystalline phase of the Co-Ni nanowire arrays was determined by XRD (X'Pert PRO-PANalytical) in a θ-2θ setup using $CuK_{\alpha 1}$ radiation ($\lambda = 1.54056$ Å). Individual alloy and multilayered nanowires were investigated by HR-TEM (JEM 2100). A chemical analysis of the multilayered nanowires was performed using TEM-EDS operating in scanning transmission electron microscope (STEM) mode. Layer thicknesses of the Cu and Co-Ni layers were calculated by combining measured Co-Ni/Cu bilayer thicknesses from TEM images (averaged over 10 bilayers) with chemical composition data from TEM-EDS. Layer thicknesses for sample Cu (0.2 nm) and Cu (0.8 nm) were determined by fitting the nominal layer thicknesses to the ratio of nominal and measured layer thickness of analyzed samples. To cross-check the crystalline phase of single alloy nanowires, SAED (JEM 2100) was performed.

*Application of the microdevice:* Diluted nanowire solution was dropped on a 150 µm thick glass substrate. The dried substrate was coated with a bilayer of lift-off and positive photoresist (Micro Chem LOR-3B and ma-P 1205) and the electrical contact structure was defined as shown in Figure 3(a), using a laser lithography system (Heidelberg Instruments µpg 101). This system was equipped with an optical microscope and a micrometer-step motor, which allows for optically scanning of the nanowires and manual definition of the contact structure position and orientation. Subsequently, the exposed parts of the resist (developer ma-D331, $t = 45$ s, $T = 20\,°C$) were removed until an undercut structure could be observed at the edges of the developed parts, which guarantees a clean removal of the metal layer in the lift-off process. Prior to the metallization, in-situ argon sputtering ($t = 15$ min, $p = 7.4 \cdot 10^{-3}$ Torr, $flow_{Ar} = 15$ sccm, $P = 20$ W) was performed to remove the protective $Si_2O$-layer from the nanowire at the exposed contact areas. A titanium layer ($d_{Ti} = 5$ nm) was sputtered. It serves as an adhesion promoter before a platinum layer ($d_{Pt} = 100$ nm) – or a gold or platinum-chrome layer – was applied. Finally, residual metallic parts were removed in a lift-off process (Remover 1165, 2×: $t = 15$ min, $T = 80\,°C$, intermediate and final cleaning with deionized water and purged nitrogen).

*Measurements:* In a cryostatic system (PPMS DynaCool from Quantum Design), transport measurements were performed on single nanowires embedded in a microdevice, see Figure 3(a), which consists of two resistive thermometers and one heater line. Voltages between 5 V and 14 V were applied (Agilent E3644A DC Power Supply) to the heater. The two resistive thermometers were used not only to measure the absolute temperature at both ends of the nanowire, but also to make electrical contact to the nanowire itself. The resistances of both thermometers were measured via four point measurement by standard lock-in technique ($I_{ac}$=10 µA, $f_1$ = 128 Hz, $f_2$ = 189 Hz) using the cryostats' integrated measurement devices. The thermovoltage $U_{thermo}$ was recorded by an additionally applied nanovoltmeter (Keithley 2184A) and the resistance $R$ of the nanowire was measured employing lock-in technique ($I_{ac}$=10 µA, $f$ = 189 Hz) supplied by the cryostatic system. The measurements were performed in a temperature range between 50 K and 325 K and at a nitrogen pressure of $4 \cdot 10^{-3}$ mbar to avoid convection. The external magnetic field, applied perpendicular to the nanowire axis, was changed in steps from -1 T to 1 T to achieve the magnetization saturation of the nanowire samples. $U_{thermo}$ and $R$ were each measured with the same applied temperature gradient to directly compare them at the very same temperature conditions.

**Acknowledgements:**

The authors gratefully acknowledge financial support via the German Academic Exchange Service, Spanish MINECO grant under research Project No. MAT2013-48054-C2-2-R, the German Priority Program SPP 1536 "SpinCAT" funded by the Deutsche Forschungsgemeinschaft (DFG), and the excellence clusters "The Hamburg Centre for Ultrafast Imaging" funded by the DFG and "Nanospintronics" funded by the State of Hamburg. Scientific support from the University of Oviedo SCTs is also acknowledged. Work carried out in Budapest was supported by the Hungarian Scientific Research Fund under Grant No. OTKA K 104696.

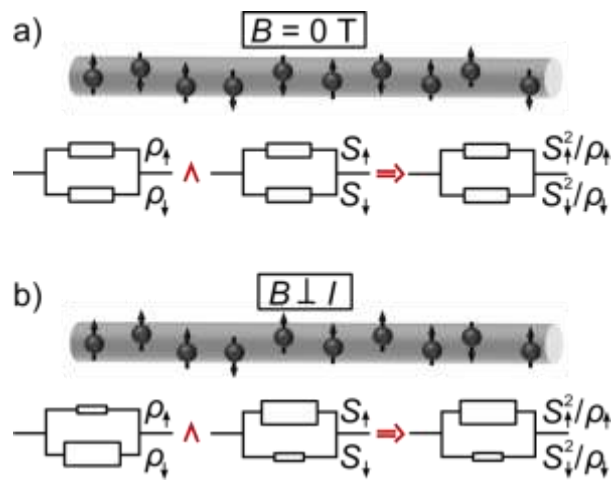

**Figure 1:** The two-current model for the resistivity as well as the thermopower $S$ and the resulting thermoelectric power factor $S^2/\rho$. a) Without an external magnetic field, the majority and minority spin channel, ↑ and ↓, exhibit equal $\rho_\uparrow$ and $\rho_\downarrow$, as well as equal $S_\uparrow$ an $S_\downarrow$, resulting in equal $S^2/\rho$ for both spin channels. b) With an externally applied magnetic field, ↑ exhibits lower $\rho$ than ↓ and according to our observation $S_\uparrow$ is increased compared to $S_\downarrow$, which results in an increase of $S_\uparrow^2/\rho_\uparrow$ and, finally, in an increase of the thermoelectric power factor of the whole system.

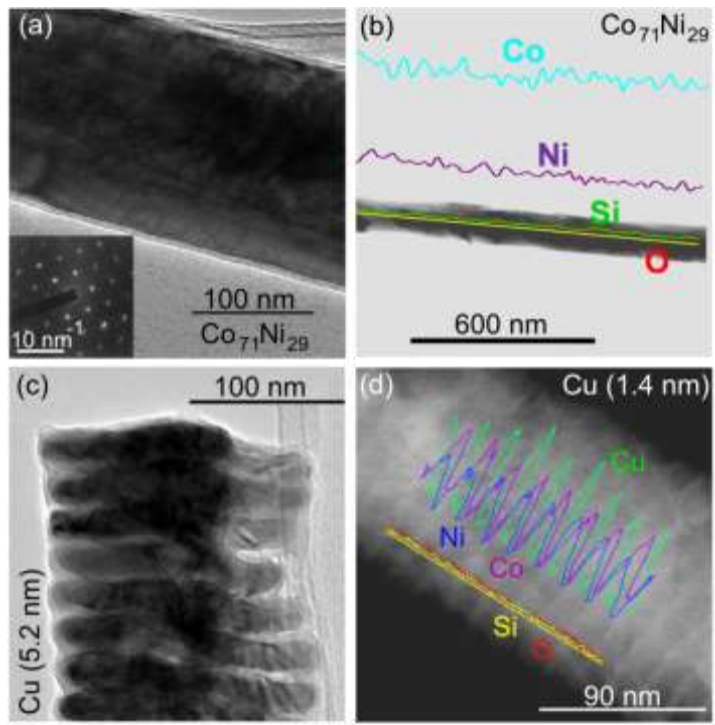

**Figure 2:** Structural, morphological, and compositional characterization of the magnetic nanowires: (a) TEM image of sample $Co_{71}Ni_{29}$. The inset shows the SAED pattern of the nanowire, which reveals the [0001] zone axis of an hcp lattice. (b) A broader TEM image of sample $Co_{71}Ni_{29}$ combined with superimposed TEM-EDS data obtained from a line scan along the nanowire length shows a fairly homogeneous Co:Ni ratio along the wire axis. (c) TEM image of sample Cu (5.2 nm) revealing the multilayered structure of the nanowire. (d) TEM image of the multilayered nanowire Cu (1.4 nm) with superimposed TEM-EDS data obtained from a line scan measured along the nanowire axis.

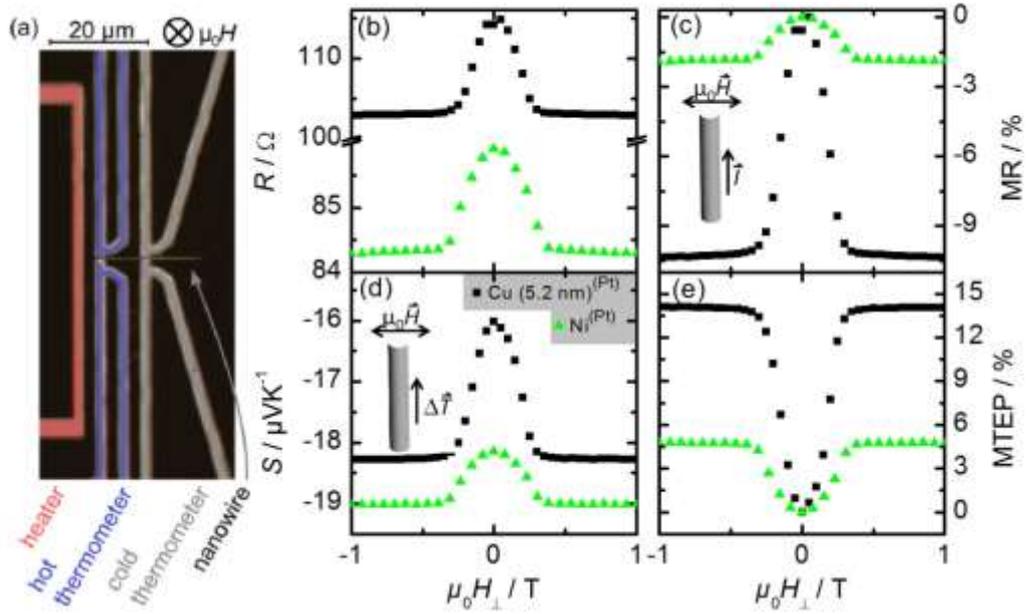

**Figure 3:** (a) SEM image displaying the inner part of the microdevice, where the nanowire is embedded between the two resistive thermometers with an additional heater line. For a nickel nanowire (green triangles) and a multilayered nanowire Cu (5.2 nm) (black squares), (b) the resistance $R$, (c) thermopower $S$ and corresponding magneto-effects (d) magnetoresistance MR (MR = $(R_H-R_0)/R_0$) and (e) magneto-thermoelectric power MTEP (MTEP = $(S_H-S_0)/S_0$) are shown as a function of the external magnetic field, $\mu_0 H_\perp$, applied perpendicular to the nanowire axis. The measurements are performed at 300 K. $S$ and MTEP of both samples are measured with respect to a platinum microdevice.

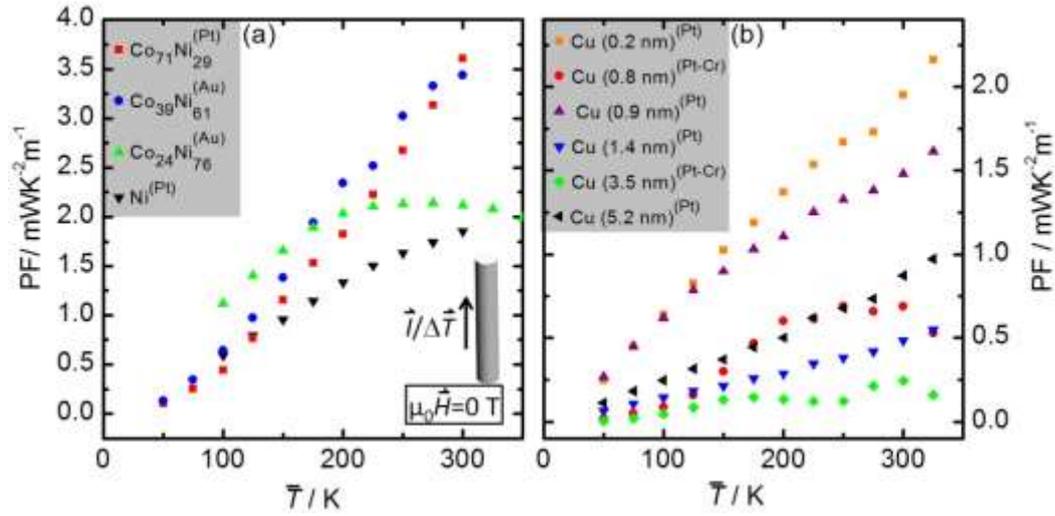

**Figure 4:** Thermoelectric power factor PF = $S^2/\rho$ of (a) single Ni and Co-Ni alloy nanowires and (b) self-standing Co-Ni/Cu multilayered nanowires measured with respect to the used electrical contact material (Pt = platinum, Au = gold and Pt-Cr = platinum-chrome) is shown as a function of the mean temperature, $\bar{T}$, of the nanowires.

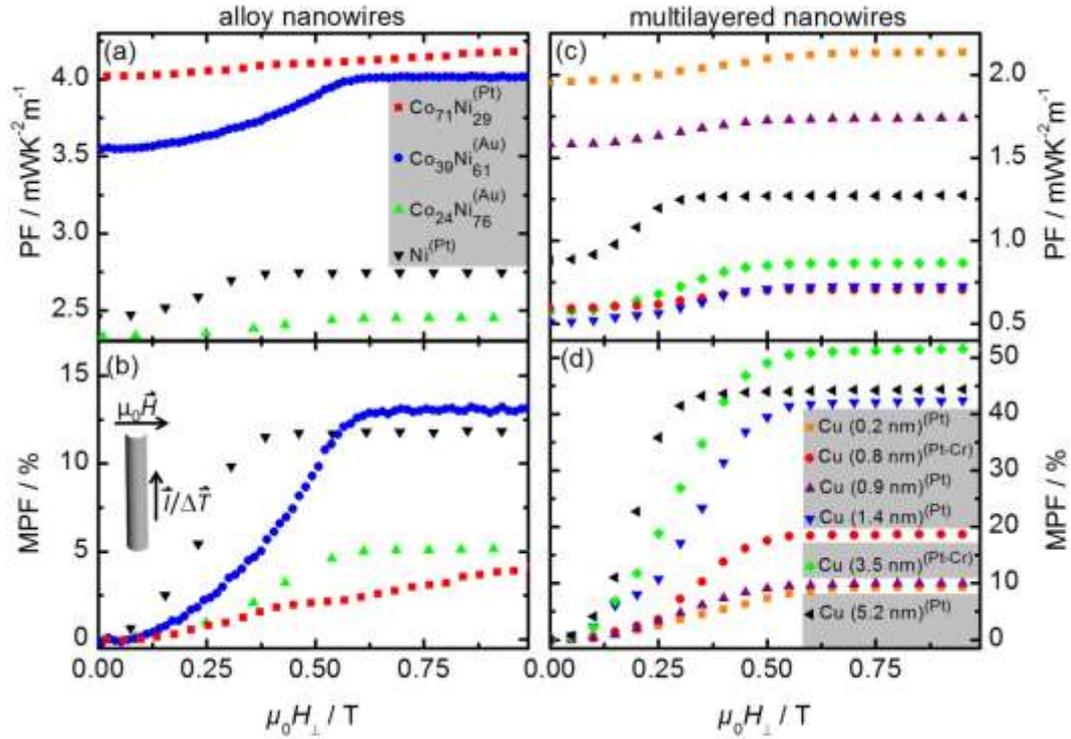

**Figure 5:** (a) Thermoelectric power factor, PF, and (b) magneto-power factor, MPF, ( MPF = (PF$_H$-PF$_0$)/PF$_0$ ) as a function of the external magnetic field, $\mu_0H_\perp$, applied perpendicular to the single Ni and Co-Ni alloy nanowire axis. Corresponding PF and MPF values for the Co-Ni/Cu multilayered nanowires are shown in (c) and (d), respectively. The base temperature was set to 300 K. Heating powers of 4.1 mW for the alloy samples and 2 mW for the multilayered samples were applied, which resulted in quite different mean temperatures $\bar{T}$ ($\bar{T} = (T_{hot} + T_{cold})/2$) of the nanowires [Co$_{71}$Ni$_{29}$: 320 K, Co$_{39}$Ni$_{61}$: 314 K, Co$_{24}$Ni$_{76}$: 342 K, Ni: 362 K, Cu (0.2 nm): 307 K, Cu (0.8 nm): 315 K, Cu (0.9 nm): 314 K, Cu (1.4 nm): 310 K, Cu (3.5 nm): 311 K, Cu (5.2 nm): 305 K] depending on the type of material and the thickness of the microdevice.

| | (a) alloy samples | | |
|---|---|---|---|
| sample | $r$(Co:Ni) % | $d$ nm | material of the microdevice |
| Co$_{71}$Ni$_{29}$ | 71:29 | 140 | Platinum |
| Co$_{39}$Ni$_{61}$ | 39:61 | 127 | Gold |
| Co$_{24}$Ni$_{76}$ | 24:76 | 140 | Gold |
| Ni | 0:100 | 126 | Platinum |

| | | (b) multilayered samples | | | | |
|---|---|---|---|---|---|---|
| sample | $\ell_{\text{bilayer}}$ nm | $r$(Co:Ni:Cu) % | $\ell_{\text{Co-Ni}}$ nm | $\ell_{\text{Cu}}$ nm | $d$ nm | material of the microdevice |
| Cu (0.2 nm) | | not measured | | 0.2 | 208 | platinum |
| Cu (0.8 nm) | | | | 0.8 | 172 | platinum-chrome |
| Cu (0.9 nm) | 17.3 ± 1.3 | 32:64:3 | 16.4 | 0.9 | 257 | platinum |
| Cu (1.4 nm) | 17.5 ± 1.5 | 47:47:6 | 16.1 | 1.4 | 205 | platinum |
| Cu (3.5 nm) | 8.7 ± 1.0 | 25:43:41 | 5.2 | 3.5 | 155 | platinum-chrome |
| Cu (5.2 nm) | 22.6 ± 1.1 | 23:54:22 | 17.4 | 5.2 | 159 | platinum |

**Table 1:** Geometrical and compositional parameters of alloy and multilayered nanowires: (a) For alloy nanowires, the atomic ratio of cobalt and nickel atoms $r$(Co:Ni) determined by SEM- and TEM-EDS, the diameter, $d$, of the nanowire as determined from scanning electron micrographs and the electrical contact material of the microdevice is given for each sample. (b) The multilayered nanowires were characterized by their bilayer thickness, $\ell_{\text{bilayer}}$, determined from transmission electron microscopy images and the atomic ratio of cobalt, nickel and copper atoms $r$(Co:Ni:Cu) measured with TEM-EDS is given. The layer thickness of the magnetic Co-Ni layer, $\ell_{\text{Co-Ni}}$, and the non-magnetic Cu spacer layer, $\ell_{\text{Cu}}$, are calculated. Also, the diameter, $d$, of the nanowire and material of the microdevice are given.